# Excited State Photophysics of Squaraine Dyes for Photovoltaic Applications: an Alternative Deactivation Scenario

G.M. Paternò,[a*] N. Barbero,[b] S. Galliano,[c] C. Barolo,[b,c] G. Lanzani,[a,d] F. Scotognella,[a,d] and R. Borrelli[e]

[a]Istituto Italiano di Tecnologia, Center for Nano Science and Technology @Polimi, Via Pascoli 70/3, 20133 Milano, Italy

[b]University of Torino, Department of Chemistry, NIS-Interdepartmental Centre, and INSTM Reference Centre , P. Giuria 7,  10125 Torino, Italy

[c]University of Torino, ICxT Interdepartmental Centre, Lungo Dora Siena 100, 10100 Torino, Italy

[d]Politecnico di Milano, Dipartimento di Fisica, Piazza Leonardo da Vinci 32, 20133 Milano, Italy

[e]Università di Torino Dipartimento di Scienze Agrarie Forestali e Alimentari, Grugliasco, I-10095

## Abstract

Squaraine dyes (SQs) represent a versatile class of functional molecules with strong absorption and emission features, widely used as near-infrared sensitizers in organic and hybrid photovoltaic devices. In this context, the photodynamics of such molecules has been seen to influence dramatically the efficiency of the photogeneration process. The most accepted interpretation of excited state deactivation in SQs is represented by a trans-cis photoisomerization around a CC double bond of the polymethinic-like bridge, although such scenario does not explain satisfyingly the decay route of SQs dyes in conformational constrained systems or in highly viscous environments. Here we combine steady-state and time-resolved spectroscopic techniques with high level *ab initio* calculations to shed light into the photophysics of *cis*-locked indolenine-based SQs. Our results point towards alternative deactivation routes, possibly involving a dark state in molecules lacking central substitution and the rotation of the central substituent in the core-functionalized ones. These novel results can suggest a synthetic rationale to design dyes that permit quantitative and effective charge generation/diffusion and collection in photovoltaic diodes and, thus, enhance their efficiency.

# Introduction

Squaraine dyes (SQs) are a class of functional molecules that exhibit a typical Donor-Acceptor-Donor molecular architecture, containing a central electron-deficient four-membered ring and two electron-donating groups. [1] The strong intramolecular charge transfer (CT) character of the $S_0 \rightarrow S_1$ transition coupled with the delocalised and extended π-electrons system, lead to strong light absorption/emission features of SQs (ε > 200000 $M^{-1}$ $cm^{-1}$)[2, 3] in the visible-near infrared (NIR) part of the electromagnetic spectrum.[4, 5] These advantageous properties alongside their thermal/chemical stability and the possibility to cover a large range of absorbed/emitted wavelengths via the versatility of chemical synthesis, have made these molecules extremely appealing for a number of applications, such as photodynamic therapy,[6, 7] biolabeling[8] and non-linear optics. [9] Furthermore, SQs have been extensively employed as NIR-sensitizers for organic (OPVs)[10] and dye-sensitized solar cells (DSCs)[11-16] and light harvesting/hole transporting layers in methylammonium lead halides perovskite photovoltaic diodes.[17] For all these reasons, it is thus important to investigate the photophysics of such functional and widely used dyes, to shed light on the excitation/deactivation mechanism occurring upon light-SQs interaction, which in turn underpins the working principles and the efficiencies of the relevant devices.

To boost further the spectral sensitivity of such dyes and hence the ability to harvest photons in the low-energy part of the solar spectrum, a plethora of SQ derivatives carrying different substituent groups attached to both the donor and acceptor blocks have been synthesized in recent years[1, 14, 15, 18-21 1, 14, 15, 18-21]. Within the context of photovoltaic applications, the introduction of a variety of substituents leads not only to a more complex deactivation pattern upon light excitation,[22, 23] but also to a profound impact on the efficiency of the photoconversion process. This, in particular, is crucial for excitonic organic and hybrid solar cells (OPVs and DSCs) in which excitons are not separated into free charges at RT due to the relatively high exciton binding energy in organic materials.[24] In fact, in this class of devices the effectiveness of charge separation, diffusion and collection processes depends sensibly on the degree of order[25-30] and conformational properties of the active materials.[31-33] More specifically, it has been proved that SQ dyes can undergo *cis-trans* photoisomerization via a twisted intramolecular charge transfer state (TICT)[34, 35,27, 28], a process that provides a non-radiative deactivation mechanism of the SQs excited state, hence hindering a fast and quantitative electron injection into the semiconductor oxide in DSCs. [35, 36,28, 29] Therefore, the inhibition of this energy-loss mechanism would enhance the efficiency of such class of devices, as it has been observed for stilbene sensitizers exhibiting a *trans-cis* photoisomerization pattern.[37] Two main synthetic strategies have been adopted to increase the population of isomers with a locked and stable *cis* conformation and thus limit the occurrence of such process: i) the insertion of a carboxy group and long alkyl chains into both the donor blocks of indolenine-based SQs; [14] ii) the introduction of bulky electron-withdrawing groups into the squaric core[3, 15, 38, 3, 15, 37]. Although both approaches prevent the intramolecular rotation owing to the increased steric hindrance of the substituent groups, the modification of the squaric central core has also a clear impact on the electrochemical, optical and structural features of SQs.[18] For instance, the introduction of a strong electron-withdrawing dicyanovinylene group into the squaric core has been observed to be beneficial for the operation of DSCs because of the improved absorption in the NIR and high-energy (≈ 400 - 450 nm) regions of the spectrum, which in turn is due to the lowering of the LUMO level and the steric hindrance offered by the bulky substituent that tends to planarize the conformation and extend the delocalization of the π-electrons system [3, 15, 18, 31, 3, 15, 18, 30] However, despite the large number of works dealing with the complex photophysics of substituted SQs[22, 23, 34, 22, 23, 33], the deactivation pattern of core-substituted dyes presenting a locked *cis* conformation has been poorly investigated, to the best of our knowledge. This is surprising, since such conformationally-frozen dyes are in fact ideal test-samples to gain insights into possible photodynamic routes different from the photoisomerization mechanism.

Here, we present a novel spectroscopic and computational study of alternative relaxation mechanisms of excited states in three indolenine-based SQ dyes carrying two octyl chains attached to both indolenine units and with different squaric ring functionalization features: the core-unsubstituted VG1-C8[14, 19], the dicyanovinylene functionalized VG2-C8[15] and the newly synthesized VG4-C8 that carries a bulky indandione group. Our spectroscopic data suggest that the key synthetic strategy limiting the photoisomerization process is the introduction of the sterically demanding alkyl chain. On the other hand, interestingly, the substitution on the central squaric ring seems to modulate the planarity of the molecule and, thus, the degree of delocalization of the π-system and the deactivation rate. Finally, high-level *ab initio* calculations suggest, for the first time, the occurrence of deactivation patterns involving a dark state and the rotation of the central substituent, which are alternative to the photoisomerization process in such conformational-locked SQs.

# Experimental Section

## Synthesis

*General remarks*

All the chemicals were purchased from Sigma Aldrich, except otherwise stated. The 2,3,3-trimethyl-3H-indole-5-carboxylic acid was supplied by Intatrade Chemicals GmbH, and was used without any further purification. All microwave reactions were performed in a single-mode Biotage Initiator 2.5. TLC was performed on silica gel 60 F254 plates using DCM and methanol (90:10) as eluents. ESI-MS spectra were recorded using a LCQ Deca XPThermo Advantage Max plus spectrometer (Thermo Finnigan), with electrospray interface and ion trap as mass analyzer. The flow injection effluent was delivered into the ion source using nitrogen as sheath and auxiliary gas.

$^1$H NMR (600 MHz) and $^{13}$C-NMR (151 MHz) spectra were recorded on a Jeol ECZ-R 600 NMR in DMSO$_{d6}$ using the DMSO signal as a reference.

*Synthetic procedures*

5-carboxy-2,3,3-trimethyl-1-octyl-3H-indol-1-ium iodide, VG1-C8 and VG2-C8 were synthesized as previously reported in the literature[14, 15, 19, 14, 15, 19].

*Triethylammonium 3-(1,3-dioxo-1H-inden-2(3H)-ylidene)-2-hydroxy-4-oxocyclobut-1-enolate (1)*

3,4-diethoxycyclobut-3-ene-1,2-dione (0.44 ml, 3.0 mmol) and 1,3-indandione (658 mg, 4.5 mmol) were dissolved in ethanol (15 ml) at 55°C. Then, tiethylamine (TEA, 0.63 ml, 4.5 mmol) was added dropwise to the mixture which assumed a dark red coloration. It was then fluxed for 4 hours, cooled down to room temperature, ethanol was partially removed to obtain a black-brown viscous liquid which was then purified by flash chromatography (up to CHCl$_3$/CH$_3$OH = 70/30) (515 mg, yield = 50 %).

UV-Vis ($\lambda_{max}$): 382 nm in MeOH.

ESI-MS: 241.18 (M-) 505.01 (2M+Na)

$^1$H-NMR (600MHz, DMSO-d$_6$): $\delta$ = 7.53 (m, 2H), 7.43 (m, 2H), 3.10 (q, 6H), 1.17 (t, 9H).

$^{13}$C-NMR (151MHz, DMSO-d$_6$): δ = 192.10, 191.52, 173.64, 139.15, 132.73, 120.44, 103.86, 42.29, 9.17.

*VG4-C8*

5-carboxy-2,3,3-trimethyl-1-octyl-3H-indol-1-ium iodide (474 mg, 1.07 mmol), triethylammonium 3-(1,3-dioxo-1H-inden-2(3H)-ylidene)-2-hydroxy-4-oxocyclobut-1-enolate (**1**) (183 mg, 0.53 mmol) and toluene/*n*-butanol (1:1, 10 mL) were introduced in a microwave vial and heated at 160 °C for 20 min until TLC and UV showed reaction completion. After solvent evaporation, the crude product was washed in hot acetone under stirring for 1 hour and filtered to obtain VG4-C8 as a dark blue powder (184 mg, yield = 41 %).

UV-Vis (λ$_{max}$): 676 nm in MeOH

ESI-MS: 835.44 (M+)

$^1$H NMR (600MHz, DMSO-d6, ppm): δ = 8.07 (d, 2H), 8.00 (dd, 2H), 7.81 (s, 2H), 7.60 (m, 2H), 7.54 (m, 2H), 7.50 (dd, 2H), 4.23 (m, 4H), 1.78 (m, 4H), 1.75 (s, 9H), 1.46 (m, 4H), 1.32 (m, 4H), 1.21 (m, 4H), 1.14 (m, 8H), 0.73 (t, 6H).

$^{13}$C-NMR (151 MHz, DMSO-d6, ppm): δ = 190.66, 177.84, 172.26, 169.78, 169.18, 167.54, 145.84, 142.73, 140.46, 132.76, 130.81, 126.94, 123.74, 120.37, 111.25, 105.18, 96.24, 49.10, 44.65, 31.65, 29.16, 29.04, 26.53, 26.43, 22.58, 14.38.

## Pump and probe spectroscopy

For the non-degenerate pump and probe measurements, we employed an amplified Ti:sapphire laser with 4 mJ output energy, 1kHz repetition rate and a central energy of 1.59 eV (800 nm). Two pump energies were used namely, 400 nm (3.18 eV) and 650 nm (1.92 eV) excitation, with the latter excitation energy in resonance with the π → π* transition. Excitation pulses at 400 nm, with pulse duration of 150 fs were obtained via second-harmonic generation in a Barium Borate (BBO) crystal. The 650 probe pulse was generated by using a visible optical parameter amplifier (OPA). [39] Pump pulses were focused on a 200 μm spot, keeping pump fluences at ≈ 1 mJ cm$^{-2}$. As a probe pulse, we used a broadband white light super-continuum generated in a CaF$_2$ window in the probe region from 340 nm to 800 nm.

The three dyes were dissolved in ethanol (d = 0.789 mg/mL, viscosity = 1.2 mPa·s at 25 °C) with a concentration optimised to give an absorbance of ≈ 1 in a 1 mm thick cuvette (0.2 mg/mL). Given the lower solubility of the three molecules in triacetin (d = 1.16 g/mL, viscosity = 17 mPa·s at 25 °C ), we used a lower concentration in this solvent (0.02 mM) to give an absorbance of ≈ 0.8, 0.4, and 0.5 O.D. for VG1, VG2, and VG4, respectively.

**Computational details**

In the VG1, VG2 and VG4 model systems the alkyl chains on the indolenine groups have been replaced by methyl groups. The SQ7 model system represents the smallest squaraine system with a polymethine chain of seven carbon atoms.

The ground state geometry of the VG1 model has been optimized using Møller-Plesset perturbation method (MP2) with the split-valence def2-SVP basis set, [40] and the geometries of VG2 and VG4 have been obtained by density functional theory (DFT) calculations with the meta M05-2X hybrid functional and the Pople 6-31G(d,p) basis set.[41] The M05-2X functional has been adopted because, having a large fraction of Hartree−Fock exchange, it yields more accurate results than standard hybrid functionals for donor− acceptor conjugate dyes. [42-44] Energies of the excited states, as well as all the potential energy profiles, have been computed using the second order algebraic diagrammatic construction (ADC(2)) method with the triple-ζ def2-TZVP basis set.[45, 46] Minimum Energy Conical Intersections (MECI) have been located using a methodology developed by Martinez in which the conical intersection (CI) geometry is obtained without the expensive calculation of the non-adiabatic derivative couplings.[47-50] ADC(2) calculation for energy and energy gradients employing a def2-SVP basis set have been used to obtain CI geometries. In order to use the ADC(2) level of electronic structure theory a locally modified version of the CIOpt software by Martìnez has been interfaced with the Turbomole software package.[51] It is worth clarifying that we are mainly interested in exploring new decay routes rather than assessing the detailed topography of the potential energy surfaces, and ADC(2) has proven to be a reliable tool in this respect.[42-44]

In order to corroborate the results obtained with the ADC(2) methodology we have performed Coupled Cluster calculations with single and double excitations (CCSD) within the equation-of-motion approach (EOM-CCSD) for the determination of the excitation energies of the model system SQ7, see figure 3. The geometry of the electronic ground state of SQ7 has been optimized at MP2 level with triple-ζ def2-TZVP basis, imposing $C_{2h}$ symmetry.[52] EOM-CCSD calculations have been performed using the Dunning correlation consistent double zeta basis set (cc-pVDZ)[53]. All calculations have been performed with the Gaussian 09 software.[54] The results of EOM-CCSD calculations on the SQ7 system are reported in table S1.

## Results and Discussion

### Synthesis

VG1-C8, VG2-C8 and the indolenine-based intermediate 5-carboxy-2,3,3-trimethyl-1-octyl-3H-indol-1-ium iodide were synthesized by a microwave-assisted method as previously reported in the literature.[14, 15, 19] The synthesis of VG4-C8 was obtained via a microwave-assisted condensation reaction of the common indolenine-based intermediate with an indandione substituted squaric acid derivative (**1**) (Scheme 1). The characterization of all the new products by $^1$H and $^{13}$C NMR and ESI-MS can be found in the SI (Figures S1−S5).

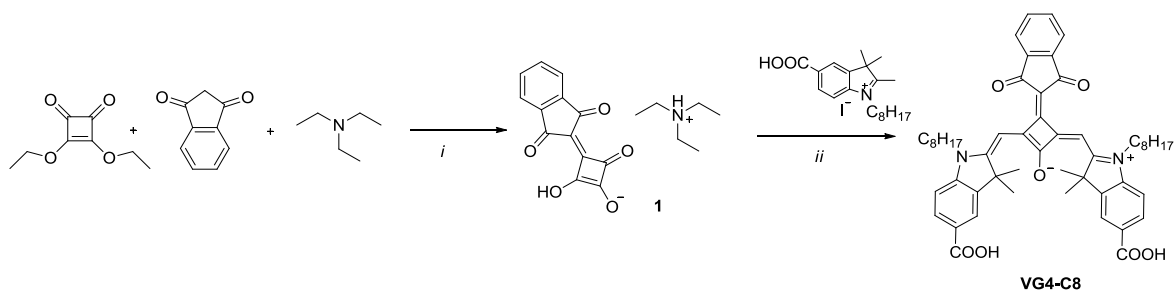

**Scheme 1.** Synthetic procedures of VG4-C8 and its intermediate. Experimental conditions: (*i*) ethanol, 4 h, reflux; (*ii*) toluene/BuOH, MW, 30 min, 160 °C.

### UV-VIS absorption and photoluminescence spectroscopies

Figure 1 shows a sketch of the molecular structures (1a), and the normalised steady-state absorption (1b) and photoluminescence (1c) spectra for the three squaraine dyes. To explore possible effects of solvent viscosity on the excitation/decaying dynamics of the dyes we dissolved the dyes in ethanol and triacetin. The three dyes exhibit cyanine-like sharp absorption bands in the visible-NIR part of the spectrum between 660-690 nm that can be assigned to the π → π* transition, whereas the shoulder located at higher energies can be interpreted as a vibronic replica.[20, 23] For the core-functionalized VG2-C8 and VG4-C8, we can observe two clear differences with the respect of VG1-C8, namely: i) an additional hypsochromic absorption band lying at 380 nm and 460 nm for VG2-C8 and VG4-C8 respectively, which can be linked to a higher-energy $S_0 \rightarrow S_2$ transition that becomes allowed due to the disruption of molecular symmetry;[3, 55] ii) a bathochromic shift of the π → π* band of VG2-C8 (690 nm) and VG4-C8 (680 nm) in respect to VG1-C8 (660 nm), which we attribute to a convolution of electronic and steric/conformational effects. In particular, the reason of such a shift might stem from both the general lowering of the LUMO level owing to the functionalization of the squaric core [18, 31],[18, 30] and to the increase of molecular planarization for substituted SQs that in turn leads to a more extended π-electrons delocalization. In this scenario, however, the replacement of the dicyanovinylene group with the more bulky indandione unit of VG4-C8 leads to an increased distortion of the π-system, reducing the π-electron delocalization and the bathochromic shift magnitude. If we pass from ethanol to the more viscous and less polar triacetin, SQs exhibit a further red-shift that follows the series VG1-C8 (9 nm) < VG2-C8 (15 nm) < VG4-C8 (20 nm). First of all, we can note that central substituted SQs display the strongest red-shift that can be ascribed to an increased dipolar character of the transition for such dyes and, in particular, to a destabilization of the HOMO level upon decrease of the polarity (solvatochromic effect[56]). Furthermore, we observe that VG4-C8 features the highest shift among the dyes, with its absorption peak almost overlapping the VG2-C8 one in triacetin. This behaviour can be connected to a mitigation of the conformational distortion of VG4-C8 in the viscous medium, corroborating the abovementioned scenario.

Photoluminescence spectra (figure 1c) allows us to corroborate the UV-VIS data, with the core-substituted SQs featuring higher Stokes shift than the unsubstituted VG1-C8 (11 nm, 19 nm and 21 nm for, VG1-C8, VG2-C8 and VG4-C8, respectively), due to the presence of the electron-withdrawing groups and the subsequent enhanced electronic asymmetry between the ground and excited states. In addition, interestingly, the slightly larger Stokes shift value of VG4-C8 also suggests that for such dye the presence of the sterically demanding indandione would introduce a distortion in the molecular planarity, which accounts for an extra contribution to the Stokes shift.[18] In the viscous and relatively apolar triacetin, the electronic and conformational asymmetry result inhibited, leading to a decrease of the Stokes shifts (7 nm, 15 nm and 15 nm for VG1-C8, VG2-C8 and VG4-C8, respectively).

To summarize this section, our findings suggest that the UV-VIS abs and PL spectra underlie a combination of electronic and conformational effects. The introduction of the electron-withdrawing groups produces an asymmetry between the ground and excited states of SQs that can explain the higher Stokes Shift and the more evident solvatochromic effect for VG2-C8 and VG4-C8. However, there are some interesting differences between VG2-C8 and VG4-C8 that cannot be explained entirely with the electron-withdrawing strength of the substituent, and which we attribute to the distortion of the molecular planarity due to the increased sterical demand of the indandione group of VG4-C8.

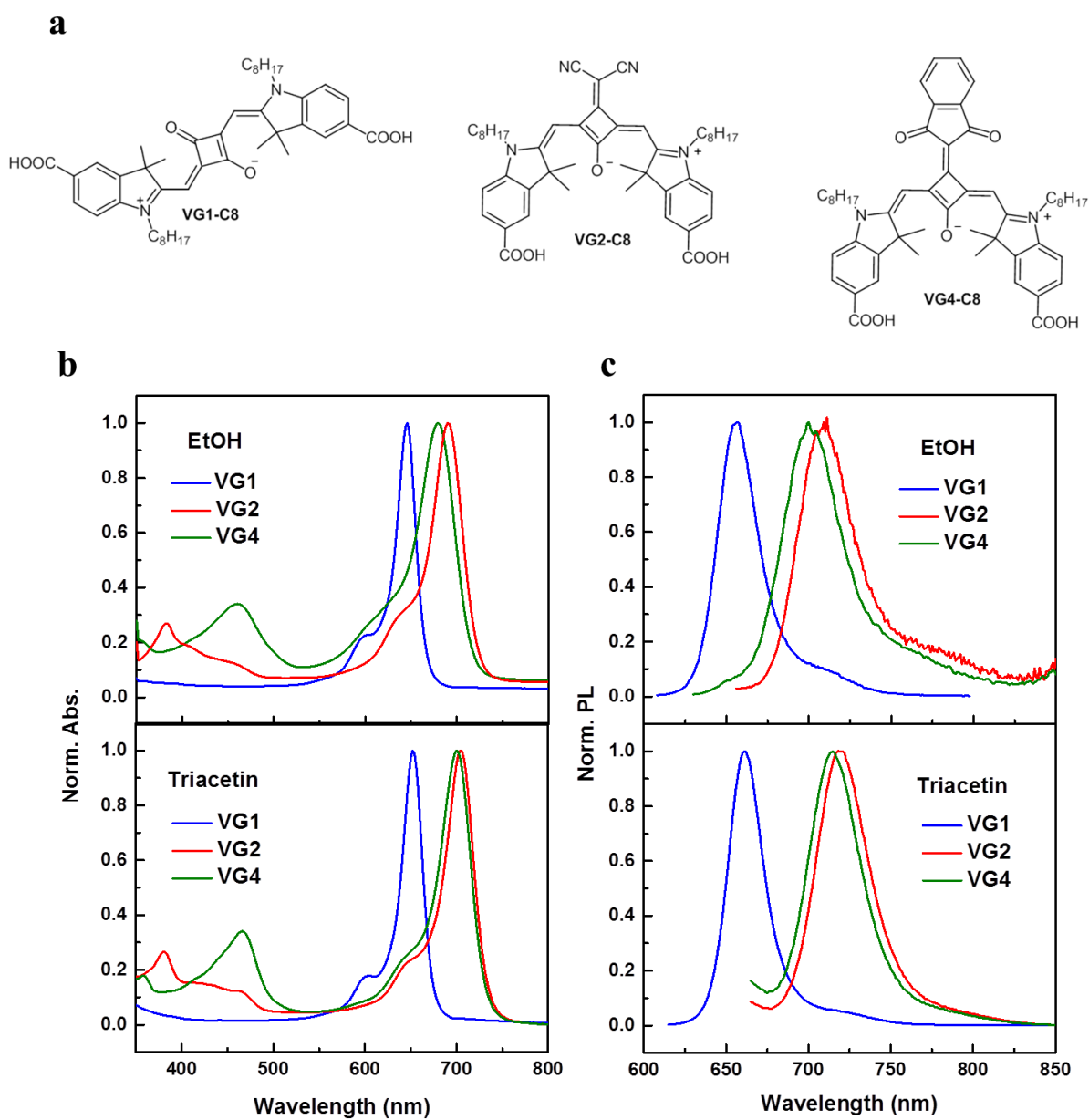

**Figure 1.** (a) molecular structures of VG1-C8, VG2-C8 and VG4-C8. (b) UV-VIS absorption and (c) Pholuminescence spectra for the three dyes in ethanol and triacetin

## Ultrafast transient absorption spectroscopy

In order to elucidate the role of the interplay between the electronic and steric properties of the side-group in determining the optical features of the SQ dyes, we performed ultrafast transient absorption in the femto- to nanosecond time-domain. The samples were excited both in resonance with the π → π* transition at 650 nm, and at 400 nm. In particular, the latter excitation energy would give a deeper insight into the effects of the conformational changes on the deactivation pathways of the molecules, since at this pump excitation more energy is distributed into the vibrational levels.

The full transient spectra for the three dyes pumped at 650 nm and 400 nm (see figure S6) report the typical transient features of symmetrical squaraine dyes and namely, i) a large positive signal in the 650-700 nm range, which can be attributed to an increase of the transmission in the sample at this wavelength as a result of the π → π* transition (photobleaching, PB); ii) a negative transient signal centred at ~ 500 nm that can be ascribed to the singlet excited state absorption $S_1 \rightarrow S_n$ (photoinduced absorption, PA). In addition, the absorption features centred at 380 nm and 450 nm for VG2-C8 and VG4-C8 respectively, are also reported as PB signal in the transient spectrum. Figure 2 shows the spectra (2a) and time-decay profiles (2b) of the PA signal of the three dyes pumped at 650 nm in ethanol and triacetin. Note that we selected the PA region to investigate the change in the transient profile at both excitation energies, to minimize any instrumental artefact and/or signal contamination coming from the pump lines. The transient spectra upon excitation at 650 nm show a general narrowing of the PA peaks passing from ethanol to triacetin, which we relate to a partially damped vibrational energy re-distribution occurring in the less viscous solvent. Yet, the red-shift of the VG1 PA in triacetin (12 nm) might be ascribed to a different stabilisation of the excited state of the dye, likely connected to its stronger ability to form H-bonds with ethanol than VG2 and VG4, i.e. involving the oxygen atom attached to the squaric ring. The time-decay profiles following the photoexcitation at 650 nm show a very long-living profile ( > 1 ns) and can be fitted using a monoexponential decay function (Table 1) in agreement with fluorescence lifetime fits (see figure S7 and Table S2 for the PL time-resolved profiles and estimated lifetimes), suggesting that the photoisomerization mechanism is already inhibited upon the introduction of the octyl chains. We observe two main effects: i) VG2-C8 exhibits the slowest dynamics among the three dyes both in ethanol and triacetin, whereas VG1-C8 and VG4-C8 show a similar profile in ethanol; ii) a slight retardation of the dynamics in triacetin for VG2-C8 (10%)  and a more substantial for

VG4-C8 (20%), while VG1 dynamics does not change appreciably upon solvent change. We explain the shorter lifetime displayed by VG4-C8 compared to VG2-C8 as a consequence of the distortion of the planarity and, thus, of the decrease of delocalization brought by the indandione group.

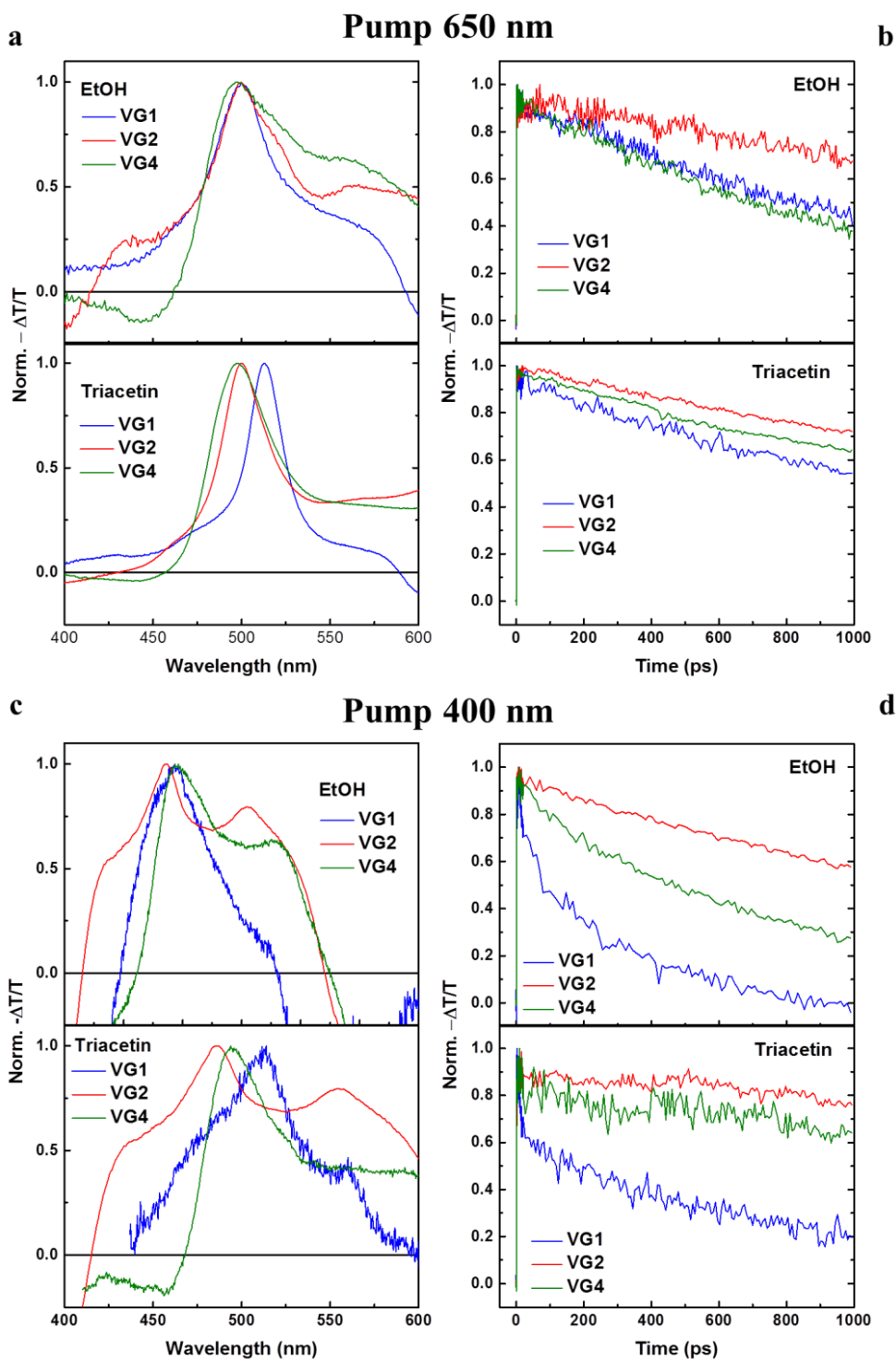

**Figure 2.** (a,c) Transient spectra of the photoinduced absorption region at 1 ps pump-probe delay of the SQs obtained by pumping at 650 nm (a) and 400 nm (c) in ethanol and triacetin. (b,d) Time-decay profiles of the dyes at a probe wavelength of 500 nm obtained by pumping at 650 nm (b) and 400 nm (d). Note that the transient signal are presented as positive values to better visualize the transient spectra and time-decay profiles.

**Table 1.** Decay lifetimes of the three SQs at pump wavelength of 650 nm and 400 nm in ethanol (black) and triacetin (red). Note the any lifetime beyond our probed time window (1 ns) should be regarded as an estimation.

| black=EtOH ; red=Triacetin | Pump 650 nm | Pump 400 nm |
|---|---|---|
| VG1-C8 | $\tau$= 1.3 ns ; $\tau$= 1.7 ns | $\tau_1$= 9.8 ps  $\tau_2$= 0.3 ns; $\tau_1$= 11 ps  $\tau_2$= 0.6 ns |
| VG2-C8 | $\tau$= 3.1 ns ; $\tau$= 3 ns | $\tau$= 2 ns ; $\tau$= 6.7 ns |
| VG4-C8 | $\tau$= 1.1 ns ; $\tau$= 2.3 ns | $\tau$= 0.6 ns ; $\tau$= 5.2 ns |

Let us now discuss the transient spectra and dynamics of the SQs upon excitation at 400 nm (figure 2c and 2d). The red-shift of the PA peak in VG1-C8, and the general lines narrowing for all the dyes passing from ethanol to triacetin are also observed at this pump energy. By pumping at 400 nm highly excited vibrational states will be populated and, as a consequence, the time-decay profiles appear modulated by the vibrational relaxation of the molecules. However, whereas for VG1-C8 the time-decay profile can be reconstructed with a bi-exponential decay function with time-constants of $\tau_1$ = 10 ps and $\tau_2$ = 300 ps, the two core-functionalized dyes exhibit a long-living mono-exponential profile with time constants of 2 ns (estimated) and 570 ps for VG2-C8 and VG4-C8, respectively . The short time-constant $\tau_1$ observed for VG1-C8 cannot be compatible neither with the rotational relaxation time (≈ 125 ps) nor with a fast intramolecular vibrational energy redistribution mechanism (IVR ≈ 70 fs) found by de Miguel et al.. [34] Here we note that the rate of such component is weakly decreased upon increase of viscosity (from 9.8 ps, weight = 50% to 11 ps, weight = 40%), implying that this deactivation route does not require the motion of sterically demanding groups, i.e. as it happens for vibrational energy cooling relaxation. [57] We anticipate that the *ab initio* calculations presented in the next section suggest that this deactivation pathway might involve a dark electronic state. Similarly to what observed by pumping at 650 nm, we note a faster decay rate exhibited by VG4-C8 than VG2-C8 that can be ascribed to its higher conformational instability likely caused by the rotation and subsequent distortion caused by the indandione group, and which seems to become milder in the viscous triacetin.

It is worth reporting that the possible structural distortion caused by the indandione group in similar SQ dyes has been studied by Mayerhöffer et al by using x-rays diffraction. [18] However our work highlights, for the first time, the direct role of central substituents on SQs deactivation route alternative to the trans-cis photoisomerization.

## Theoretical analysis

The photodynamics of squaraine dyes shows several aspects whose interpretation requires special care. The first observation is that the excited state lifetime of VG1-C8 is weakly dependent on the solvent viscosity. The current interpretation is that excitation to $S_1$ state induces a *cis-trans* isomerization around a double CC bond of the polymethinic-like bridge with a very small yield, less than 1%.[58] Embedding in a rigid polymer matrix seems not to inhibit the decay, leading to the conclusion that the molecules are not tightly surrounded by the polymer.[35] In order to rationalize this puzzling behaviour, we have investigated the topography of the potential energy surfaces of several electronic states in model squaraine molecules by means of high level *ab initio* electronic structure calculations (see computational details).

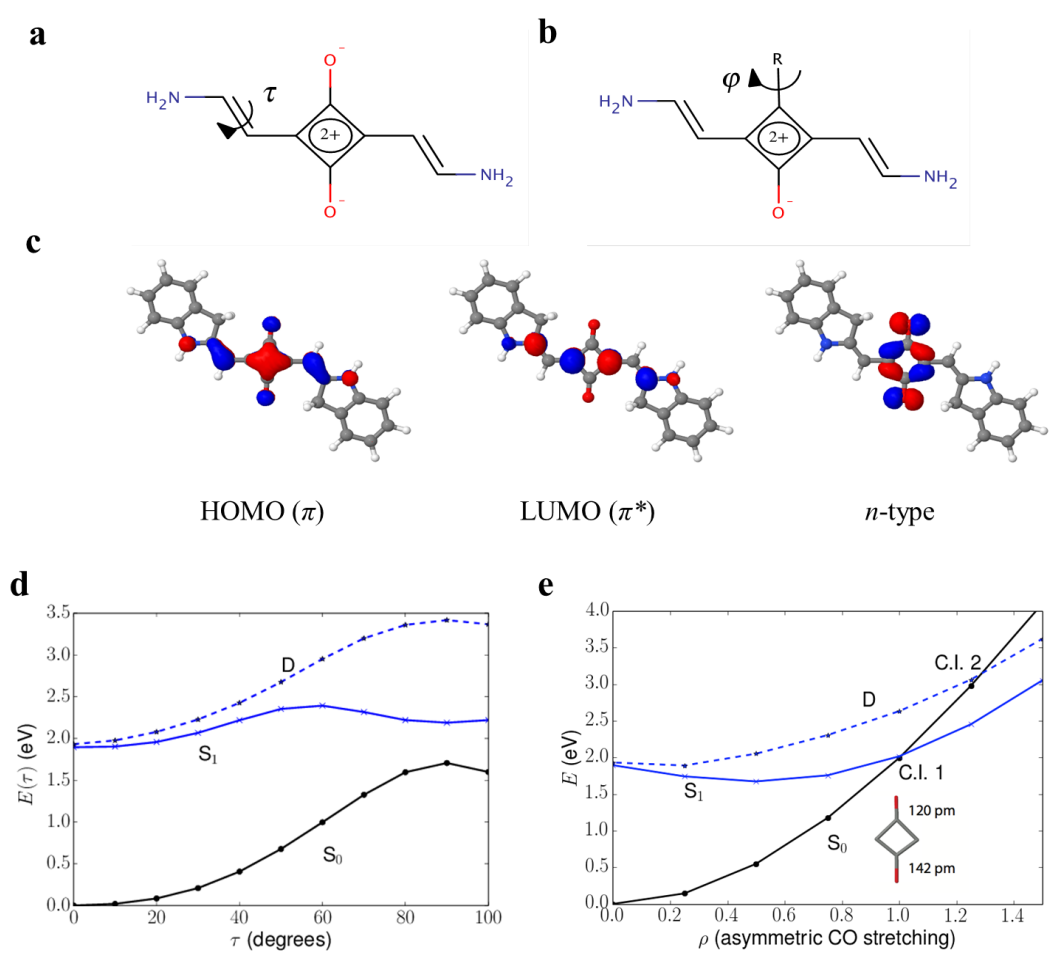

**Figure 3.** (a) Definition of the torsional coordinates t and j in squaraines and (b) centrally substituted squaraines. The molecule (a) represents the SQ7 model used in CCSD calculations. (c) Molecular orbitals of VG1. The HOMO and LUMO orbitals have a π character and is delocalized over the polymethine chain, whereas the n-type orbital involves exclusively the squaric subunit. (d) Computed potential energy curve for the cis-trans isomerization of a model VG1 system. (e) Computed potential energy curves along the asymmetric CO stretching coordinate (see supplementary information for the definition of the coordinate).

Figures 3d shows the computed potential energy profiles along the isomerization coordinate τ, as defined in figure 3a for a model VG1 system (see computational details). We immediately notice that in the Franck-Condon region the first two excited electronic states of VG1 are very close in energy, but only one is bright. We will refer to the bright state as $S_1$, and to the dark state as D. The former is obtained from a HOMO-LUMO (π-π*) excitation involving orbitals of the polymethine chains, while the latter results from an n-π* type excitation which involves the lone pair of carbonyl groups and the π* orbital of the polymethine chain (figure 3c). The energy gap between the ground and first excited state lowers as τ approaches 90°, reaching a value of about 0.45 eV at the local energy minimum of $S_1$. The geometry of the minimum energy conical intersection point (MECI) between the $S_0$ and $S_1$ electronic states is very close to the twisted structure, and its energy is reported on the same figure. This confirms that the torsional motion of the indolenine groups can provide a direct deactivation path of the $S_1$ excited state. Furthermore, as is evident from the computed energy profiles, the dark state is not directly involved in this route since the energy gap with the other electronic states increases along the torsional coordinates.

On the other hand, the quasi degeneracy of the D and $S_1$ states at the FC point, suggests that the D state might be effectively involved in the decay mechanism of squaraines. In order to support this idea we have searched for alternative crossing regions that would not involve large distortion of the torsional coordinate. More specifically, using a technique for MECI localization by Martìnez, [47] we have been able to identify a new decay route, which involves primarily the asymmetric elongation of the carbonyl bonds of the squaric unit. Figure 3e shows the computed potential energy curves as a function of the asymmetric elongation of the CO bonds (see supporting information for details on the definition of the reaction coordinate). At the minimum energy point the two CO bonds have equal lengths of about 1.22 Å. As the asymmetric CO increases the energy of the D state lowers while that of $S_0$ and $S_1$ increases. Eventually the D state becomes the lowest energy state crossing $S_0$ at the C.I. point 1, where the CO bond lengths are 1.42 Å and 1.20 Å, while the potential energy curves of $S_1$ and $S_0$ crosses at the point C.I. 2, about 0.6 eV above the Franck-Condon (FC) excitation region. Since the $S_1$ and D states are almost degenerate in the FC region the transition D ← $S_1$ can be easily attained after excitation to $S_1$; from the D state the system can decay to the ground state via the C.I. 1, by a large CO stretching motion. The energy of the second conical intersection (C.I. 2) is too high to be accessible after excitation in the FC region. According to the current findings the deactivation of the first bright state of squaraines can follow multiple channels. In particular both cis-trans isomerization as well as asymmetric CO stretching, appear to be active. Interestingly, a major difference between the two processes is that the asymmetric CO stretching does not involve the motion of large and sterically demanding substituents, making it a potentially preferred route for conformational locked molecules, especially in a high viscosity environment. This, in particular, can explain the deactivation pattern exhibited by VG1-C8, for which we could not observe any appreciable dependence of the transient absorption dynamics upon increase of the viscosity. Finally, since this route involves the squaric ring it is expected to be typical of all squaraines, at least of those lacking central substitutions such as VG1-C8.

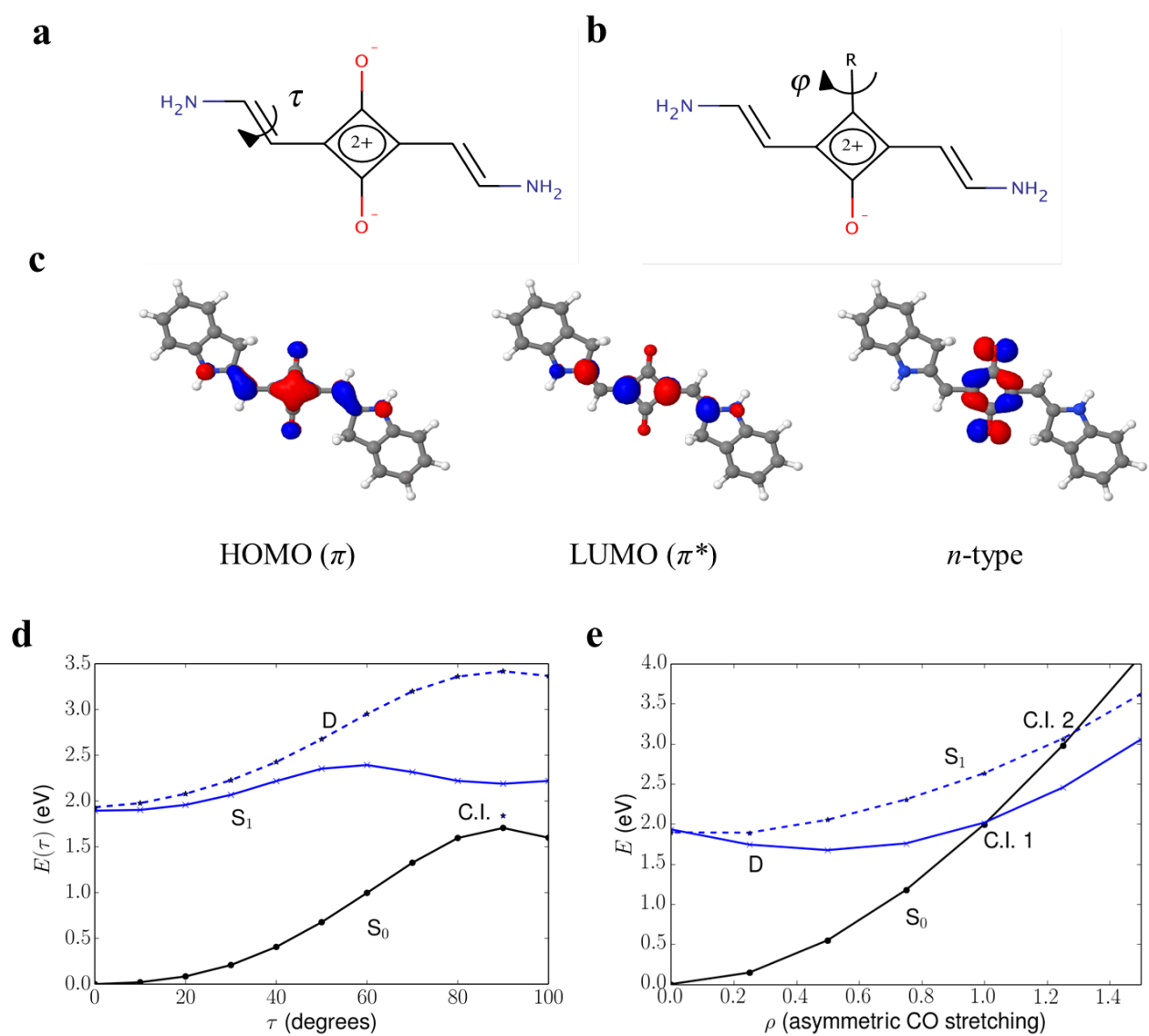

**Figure 4.** (a) Potential energy curves of the first three electronic states of VG2 computed at ADC(2)/def2-SVP along the polymethine twisting and (b) the twist of the substituent group. (c) Potential energy curves of the first three electronic states of VG4 computed at ADC(2)/def2-SV(P) along the polymethine twisting and (d) the twist of the substituent group.

To better assess the role of the electron withdrawing groups of VG2-C8 and VG4-C8 we have also studied the variation of the potential energy curves as a function of the rotation angle, $\varphi$, of the central substituent as defined in figure 4. In particular, the rotation of the dicyano group in VG2-C8 leads to a destabilization of both the ground and first excited state energies, with an overall narrowing of the $S_0/S_1$ energy gap to about 0.6 eV. The same holds for the rotation of the indandione group in VG4, as clear from figure 4c,d. This suggests that the internal rotation can have a role in the decay of the $S_1$ excited electronic state, however we have not yet investigated the existence of a real conical intersection between the two potential energy surfaces. From our results, interestingly, VG2-C8 and VG4-C8 lack the D dark state. This is very likely due to the substitution of the oxygen atom of the squaric ring with a highly electron-withdrawing group, as the n orbitals of the carbonyl are not available. On the other hand this does not rule out the possibility that the deformation of the squaric ring lead to a crossing of the ground and $S_1$ electronic states as happens in VG1-C8. Further investigations are in progress along these lines.

In this scenario, passing to the experimental results, the sterically demanding indandione group would activate such deactivation pathway, as it would increase the conformational instability of the central part, leading to an enhanced rotation of the substituent group and, hence, to a faster decay of the excited state than the other centrally substituted VG2-C8.

## Conclusions

In summary, we have reported on a spectroscopic and computational study on three indolenine-based SQ dyes carrying two octyl chains attached to both the indolenine groups and with three different central functionalization pattern: the non-substituted VG1-C8, dicyanovinylene functionalized VG2-C8 and the newly synthesized VG4-C8 with a bulky indandione group. Our findings indicate a strong inhibition of the photoisomerization mechanism already upon insertion of the bulky alkyl chain in the donor indolenine blocks. The weak dependency of the deactivation pattern of VG1-C8 upon increase of solvent viscosity suggests the involvement of an alternative excited states decay pattern, which does not rely on the motion of sterically demanding groups. High-level *ab initio* calculations identify, for the first time, the role of a dark state on the deactivation routes of VG1-C8 and, in general, of SQs dye. On the other hand, core-substituted dyes lack of this route, showing an alternative deactivation mechanism that encompasses the rotation of the substituent. In this scenario, the bulky indandione group of VG4-C8 likely leads to a more pronounced distortion of the squaric core and, thus, to a more conformational freedom than VG2-C8 and to a more efficient deactivation of the excited state. These novel findings would be of general interest for minimizing the deactivation loss in SQ dyes and, therefore, for enhancing the efficiency of related devices.

## Conflicts of interest

There are no conflicts to declare.

## Acknowledgements

R. B. acknowledges the CINECA award QDYMA-HP10CMJCGW under the ISCRA initiative, for the availability of high performance computing resources and support, and the financial support of the University of Torino through the grant BORR-RILO-16-02. This work is supported by the H2020 ETN SYNCHRONICS under grant agreement 643238.

*Supplementary Information for*

**Excited State Photophysics of Squaraine Dyes for Photovoltaic Applications: an Alternative Deactivation Scenario**

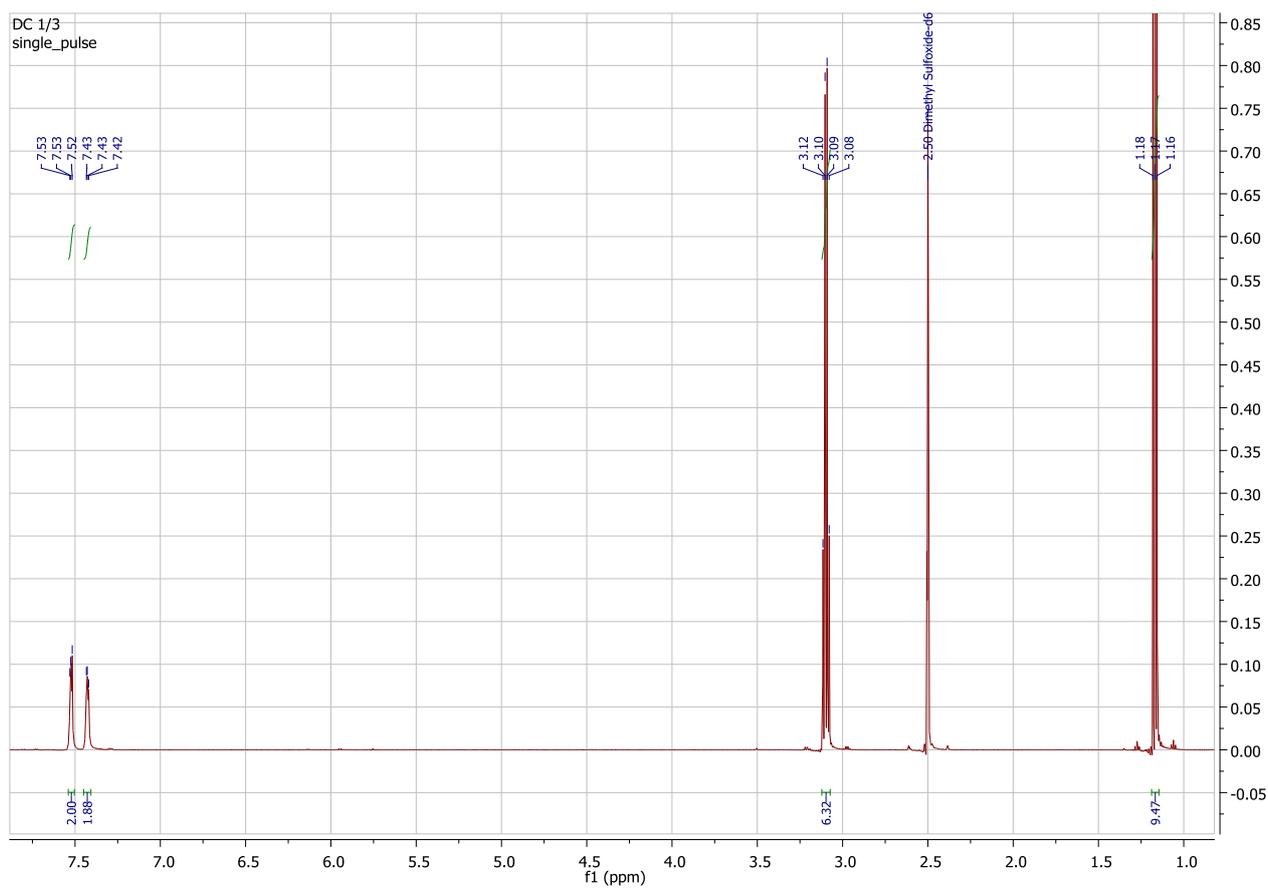

**Figure S1**. $^1$H NMR of intermediate **1**.

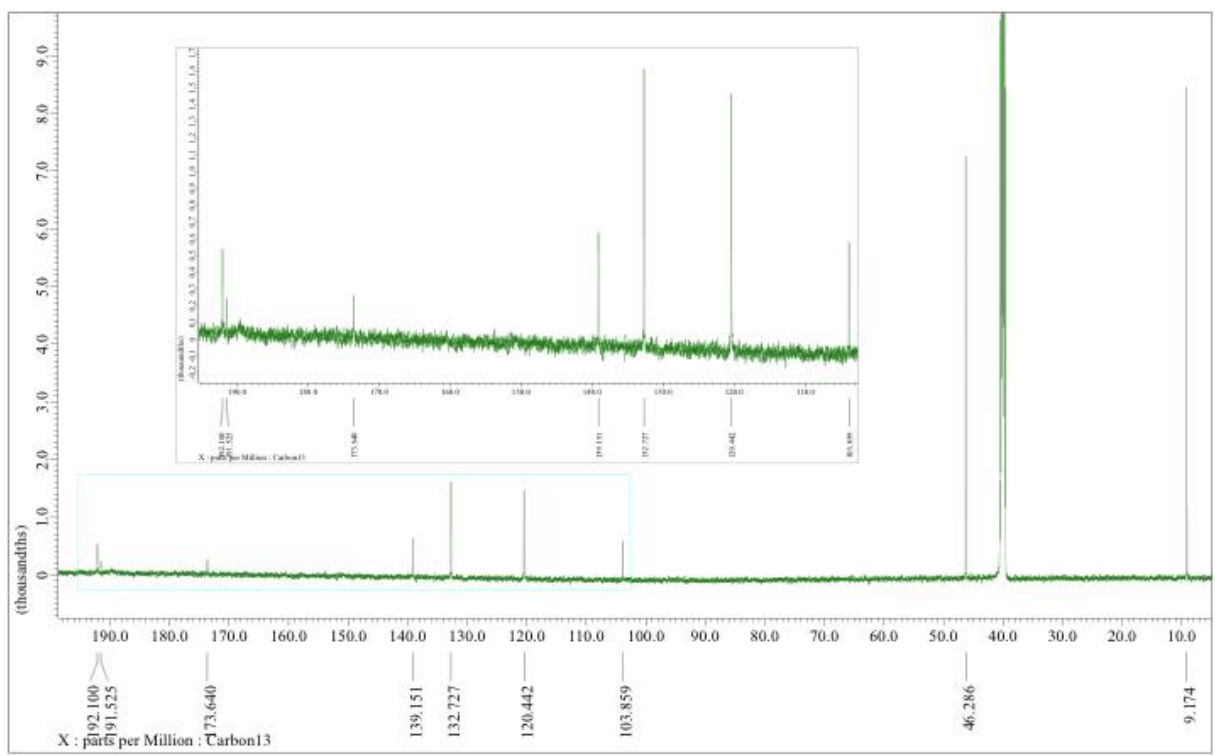

**Figure S2**. $^{13}$C NMR of intermediate **1**.

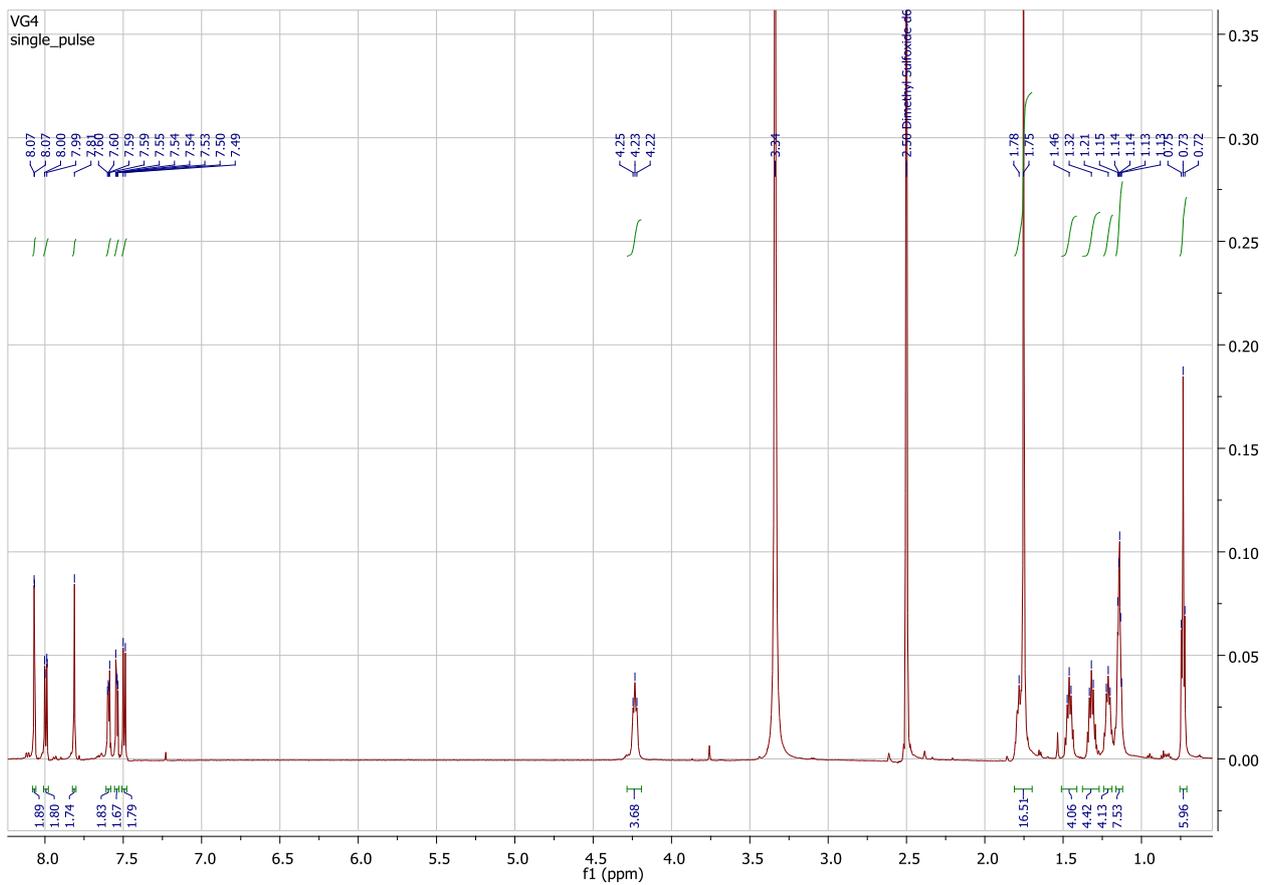

**Figure S3**. ¹H NMR of squaraine dye VG4-C8.

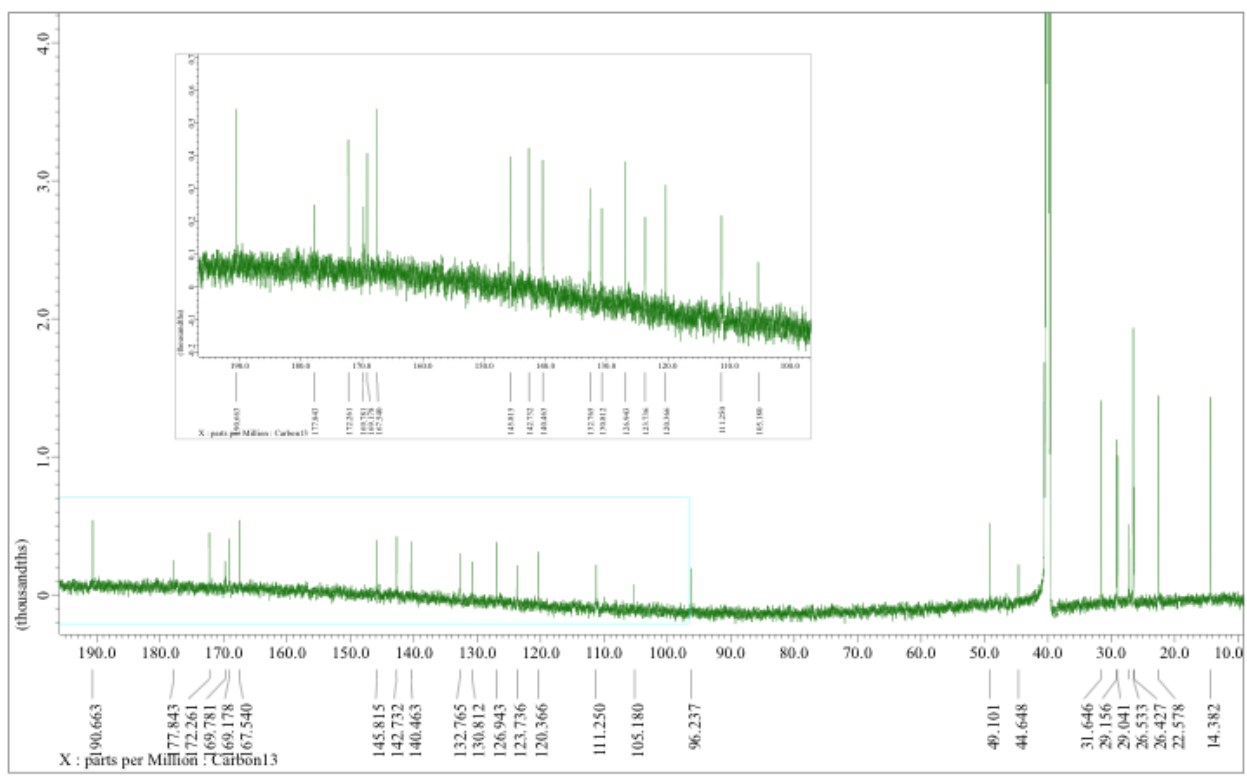

**Figure S4**. ¹³C NMR of squaraine dye VG4-C8.

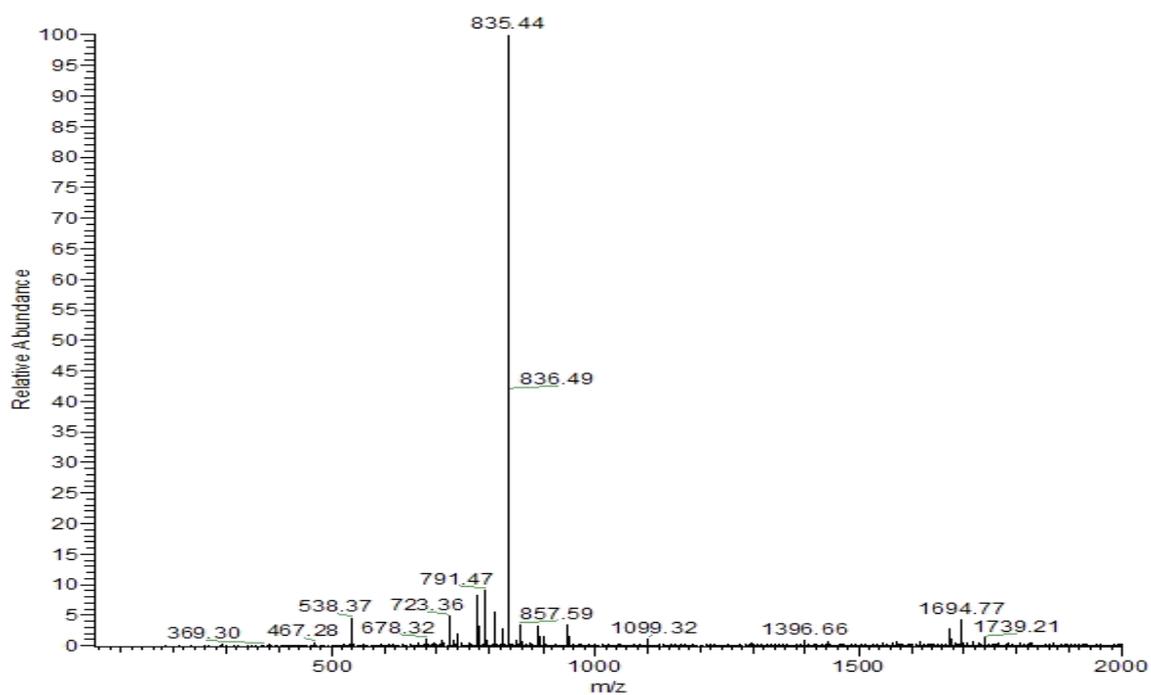

**Figure S5**. ESI-MS of squaraine dye VG4-C8.

**Table S2.** Excitations energies and oscillator strengths of SQ7 computed at EOM-CCSD level with a cc-pVDZ basis set. The transition to the D state having Bg symmetry is forbidden by symmetry in the C2h point group.

| Electronic state | Excitation energy (eV) | Oscillator strength |
|---|---:|---:|
| $S_1$ ($B_u$) | 2.9565 | 0.9676 |
| $D$ ($B_g$) | 2.9870 | 0.0 |
| $S_2$ ($B_g$) | 4.9815 | 0.0 |

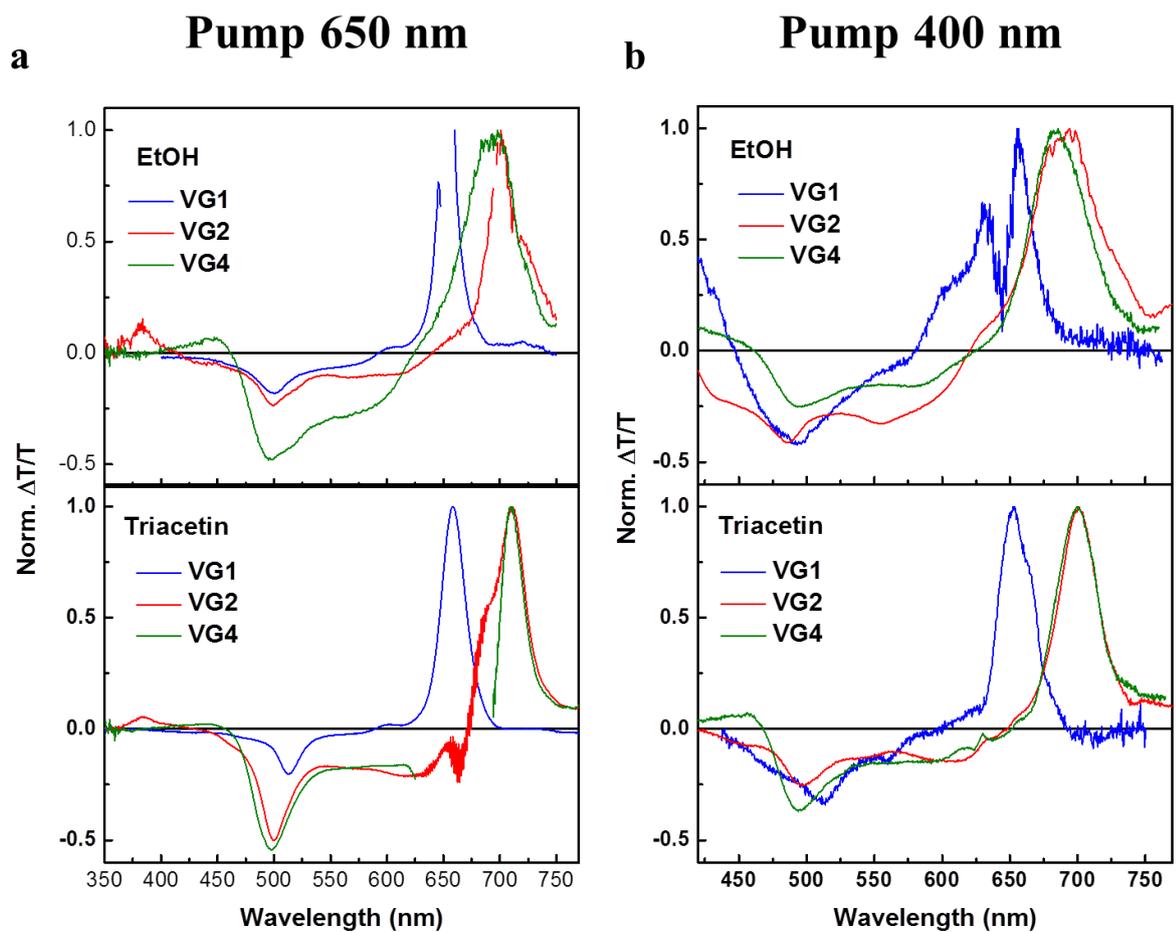

**Figure S6.** Full transient spectrum for the three dyes excited at (a) 650 nm and (b) 400 nm in ethanol and triacetin. The signal at 650 nm is contaminated by the scattering of the pump line. For this reason, we selected the photoinduced region to analyse the spectral and dynamical features of the excited states of SQs.

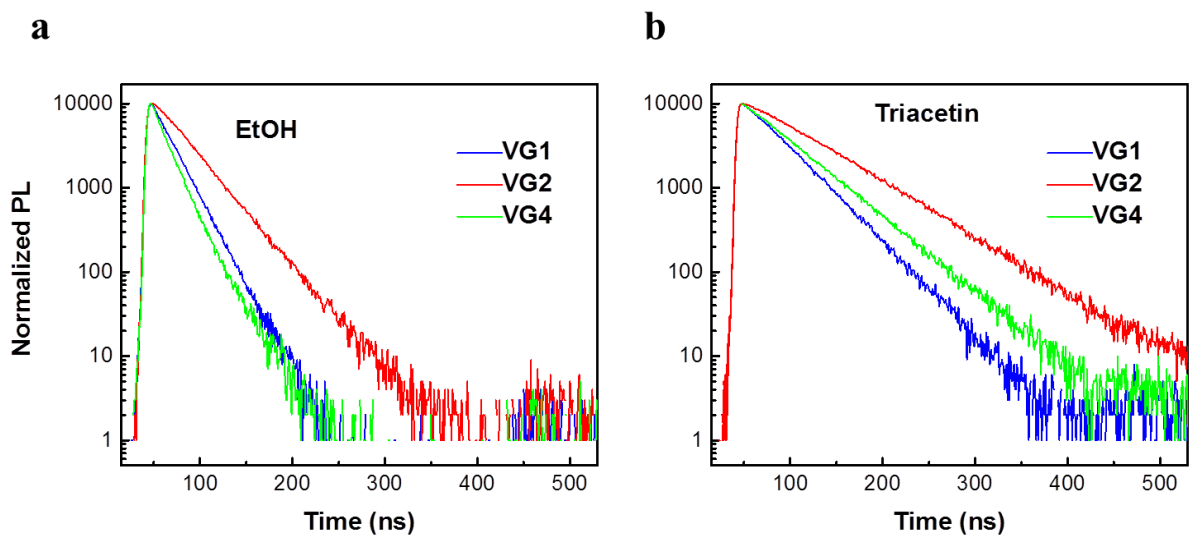

**Figure S7.** (a) Normalized time-resolved PL profiles for the three dyes in ethanol and (b) triacetin taken with an excitation wavelength of 650 nm. The y-axis is reported in logarithmic scale to highlight the monoexponential decay of the time-profiles.

**Table S3.** PL decay lifetimes of the three SQs at pump wavelength of 650 nm in ethanol and triacetin.

|  | Time-constant in ethanol (ns) | Time-constant in Triacetin (ns) |
|---|---|---|
| **VG1-C8** | 21.1 | 42.3 |
| **VG2-C8** | 35.8 | 72.5 |
| **VG4-C8** | 16.5 | 50.4 |

**Computation of the energy profile of Figure 3e**

Following the work by Martinez,[R3] the MECI point has been located by minimization of the objective function

$$F_{IJ}(\boldsymbol{R}) = E_{IJ}(\boldsymbol{R}) + G_{IJ}(\Delta E_{IJ};\boldsymbol{R})$$

where $\boldsymbol{R}$ is the column vector with the 3N Cartesian coordinates of the N atoms of the system and

$$E_{IJ} = [E_I(\boldsymbol{R}) + E_J(\boldsymbol{R})]/2$$

$$\Delta E_{IJ} = E_I(\boldsymbol{R}) - E_J(\boldsymbol{R}).$$

$G_{IJ}(\Delta E_{IJ};\boldsymbol{R})$ is a properly chosen penalty function that increases monotonically with the energy gap. Once the C.I. geometry ($\boldsymbol{R}_{CI}$) has been located we use a linear interpolation scheme to displace the atoms from the equilibrium geometry, $\boldsymbol{R}_0$, of $S_0$ to the C.I. point, and define a reaction parameter $\rho$ such that

$$\boldsymbol{R}(\rho) = \boldsymbol{R}_0 + \rho \Delta \boldsymbol{R}$$

With $\Delta \boldsymbol{R} = \boldsymbol{R}_{CI} - \boldsymbol{R}_0$. Clearly $\boldsymbol{R} = \boldsymbol{R}_0$ when $\rho=0$ and $\boldsymbol{R} = \boldsymbol{R}_{CI}$ when $\rho=1$. Intermediate geometries are obtained by letting $\rho$ vary between 0 and 1.